\documentclass[useAMS,usenatbib]{mn2e}
\usepackage{graphicx}
\usepackage{subfigure}
\usepackage{afterpage}
\usepackage{txfonts}
\usepackage{booktabs}
\usepackage{natbib}
\usepackage{paralist}

\def \kms       {\hbox{km$\,$s$^{-1}$}}

\title{GSH~90-28-17: a Possible Old Supernova Remnant}
\author[L. Xiao and M. Zhu]
       {L. Xiao$^1$\thanks{E-mail: xl@nao.cas.cn} and M. Zhu$^1$  \\
      $^1$National Astronomical Observatories, Chinese Academy of
      Sciences, Jia-20, Datun Road, Chaoyang District, Beijing 100012,
      China}
\date{Released 2013 Xxxxx XX}

\pagerange{\pageref{firstpage}--\pageref{lastpage}} \pubyear{2013}

\begin{document}

\label{firstpage}

\maketitle

\begin{abstract}
GSH~90-28-17 is a high-latitude galactic HI supershell, identified in the HI
supershell catalogs with a velocity of $v_{lsr}\sim-17$~\kms. We used the new
Arecibo GALFA-HI survey data which have much higher resolution and sensitivity
than what were previously available to re-examine the properties of the supershell.
We derived a new distance of 400~pc for GSH~90-28-17 and suggested that it is
related to the Lac OB1 association. The radius of GSH~90-28-17 is 66.0$\pm$3.5~pc.
The HI mass of the shell is (3.1$\pm0.1)\times10^{4}$~M$_{\odot}$.
It has an age of $\sim4.5$~Myr and a total kinetic energy of (8.2$\pm0.3)\times10^{48}$~ergs.
We extracted radio continuum data for the GSH 90-28-17 region from the 408~MHz
all-sky Survey and Bonn 1420~MHz survey, and filtered the diffuse background
Galactic emission. A radio loop-like ridge is found to be associated with the HI shell
at both frequencies, and shows a nonthermal origin with a TT-plot index of $\alpha$=-1.35$\pm$0.69. 
In addition, the pulsar J2307+2225 with a similar distance is found in the shell region. 
We conclude that GSH~90-28-17 is probably an old, type II supernova remnant in the solar
neighborhood.
\end{abstract}


\section{Introduction}

A large number of large-diameter HI shells (or
supershells)~\citep{h81,h84} have been found in a series of 21~cm line
surveys, e.g.~\citet{hh74} and~\citet{cph80}. It has been suggested that 
these HI supershells could be the result of some kind of explosion events
that occurred in stellar associations, and they are commonly interpreted as old
supernova remnants (SNRs) with a typical age of
$t~\sim10^{6}$~yr~\citep{h84}. As old SNRs evolve into the late
stage, the radio surface brightness declines and the interior temperature 
drops that SNRs gradually fade away. The accumulated
neutral hydrogen shells will last longer until the velocity of the
expansion decreases down to the random velocity of the interstellar
medium (ISM).Those that exploded at very high galactic latitudes are probably
able to avoid the interstellar turbulence and supernova explosions that frequently
occur in the near-plane region, and can evolve till old age.

GSH 90-28-17 is a high-latitude galactic symmetrical HI supershell
centered at (l, b)=(90$\degr$,$-28\degr$), with a velocity
$v_{lsr}=-17$~km~s$^{-1}$.  The shell has an angular extent of
approximately 18$\degr\times21\degr$ in the sky, with an expansion
velocity of about 4.5~km~s$^{-1}$. It was first completely identified
and listed in the catalog of HI supershells by~\citet{h84} from the
data of the Berkely survey~\citep{hh74} and the HI survey
of~\citet{cph80}. The distance of the shell was estimated as 
$\sim$3.8~kpc~\citep{h84} based on a flat rotation curve model for the Galaxy. 
However, as mentioned in~\citet{h81}, distance estimates based on rotational
velocity, a method often used in the low latitude region, are not suited for
shells at high latitude. ~\citet{h81} also identified a partial shell
(GS 91$-$33) of GSH 90-28-17 using a filtering method, and derived a
lower limit on the distance of 300~pc for it based on extinction versus
distance studies along the line of sight.  The kinetic energy of the
shell GS 91$-$33 is estimated to be about 10$^{48}$~ergs with an age
of 1.5$\times10^{6}$~yr~\citep{h81}.  Using the distance of 300~pc,
the diameter of GSH 90-28-17 is about 100~pc, and it is at a vertical distance
of z$\sim$140~pc from the Galactic plane.

\citet{sn83} searched for low-brightness radio continuum loops
possibly associated with HI filaments in the 408 and 820~MHz survey
maps. They applied the "background filtering" (BGF) method to remove
large-scale background emission, and found a radio continuum loop
associated with the HI shell GS 99$-$26 with a diameter of 7$\degr$,
which is in the northeast and adjacent to GSH 90-28-17.  The distance
of the radio loop is estimated to be d$\sim$570~pc using the surface
brightness-diameter ($\Sigma-D$) relation of~\citet{m79}, assuming it
is a normal SNR~\citep{sn83}. However, they excluded the physics
association between GS 99$-$26 and the radio loop by adopting a
distance of 3.5~kpc for the HI shell derived from the Galactic rotation curve.

Most supershells are found to be associated with OB associations.
In the vicinity of GSH 90-28-17 lies the famous Lac OB1 association located
at a distance of 370~pc~\citep{zhb99}. Could this supershell have a physical
connection with the Lac OB1 association {\it i.e.} caused by a progenitor
star running out of it? A detailed analysis is required before any firm
conclusion can be drawn. As a high latitude supershell far from the plane,
the lack of an absorption effect caused by intervening material makes GSH 90-28-17
a good sample to study the evolution of HI supershells under a low-density
environment. Former HI surveys did not provide enough sensitivity for the
interpretation of physical parameters of GSH 90-28-17. The newly published
GALFA HI survey~\citep{phd11} has produced high sensitivity and high resolution HI
data cube with $\sim 4\arcmin$ resolution, which enables us to clearly identify
the structure and dynamics of GSH 90-28-17. Existing multi-wavelength
observations also enable us to derive the physical properties of the HI
supershell, and study its association with radio continuum emission.

The plan for the paper is as follows. In Sect. 2 the observational HI and
radio data set we used are described. In Sect. 3 we present detailed neutral
hydrogen and radio continuum images of GSH 90-28-17 as well as analysis results. 
The distance estimation of GSH 90-28-17 and its relationship with 
the Lac OB1 association are also presented. The possible origin for energy in
GSH 90-28-17 is discussed in Sect. 4. In Section 5 we summarize our main conclusions.

\section{Observational data }
\subsection{HI data}
The HI data sets come from the recently released Galactic Arecibo L-Band Feed
Array HI survey (GALFA-HI), described in detail by~\citet{phd11}. The angular
resolution of the survey is about 4$\arcmin$.  It covers a wide velocity range
from $-$700~km~s$^{-1}$ to +700~km~s$^{-1}$ with a velocity resolution of
0.18~km~s$^{-1}$. Typical noise levels are 80~mK in an integrated 1~km~s$^{-1}$ 
channel.

\subsection{Radio data}
The 408~MHz radio continuum data used here are taken from the 408~MHz all-sky Continuum
Survey~\citep{hss82}, which is a mosaic of data taken at the Jodrell Bank MkI, Effelsberg
100~m, Parkes 64~m and Jodrell Bank mkIA telescopes. The angular resolution is
$0\fdg85$. The 1420~MHz radio continuum data are taken from the Bonn 1420~MHz survey
~\citep{r82,rr86} with the Bonn Stockert 25~m telescope. The resolution of the survey
is about 35$\arcmin$, and the effective sensitivity is about 50~mK~T$_{\rm B}$.

\section{Results}
\subsection{the HI Morphology}
\begin{figure*}
\centering
\includegraphics[angle=0,width=0.33\textwidth,bb=0 0 235 207]{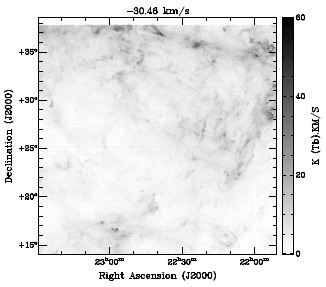}
\includegraphics[angle=0,width=0.33\textwidth,bb=0 0 238 212]{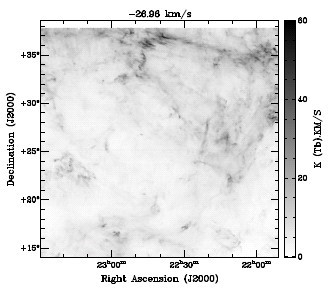}
\includegraphics[angle=0,width=0.33\textwidth,bb=0 0 240 207]{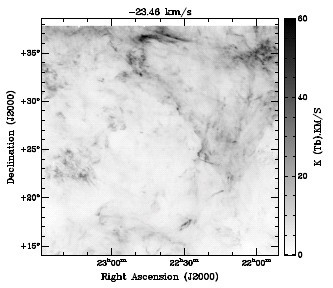}
\includegraphics[angle=0,width=0.33\textwidth,bb=0 0 237 207]{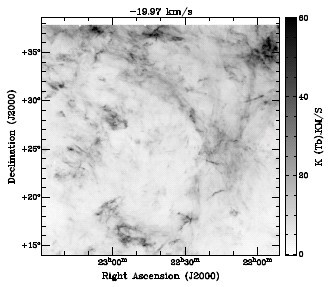}
\includegraphics[angle=0,width=0.33\textwidth,bb=0 0 236 206]{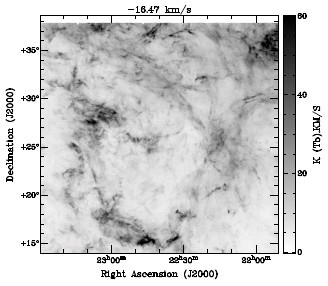}
\includegraphics[angle=0,width=0.33\textwidth,bb=0 0 239 208]{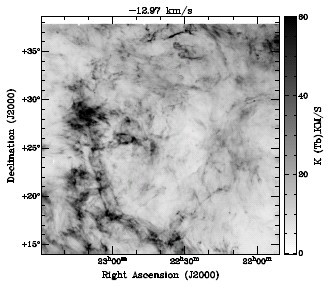}
\includegraphics[angle=0,width=0.33\textwidth,bb=0 0 239 207]{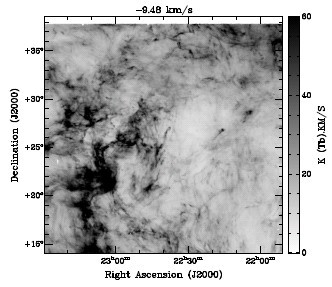}
\includegraphics[angle=0,width=0.33\textwidth,bb=0 0 239 209]{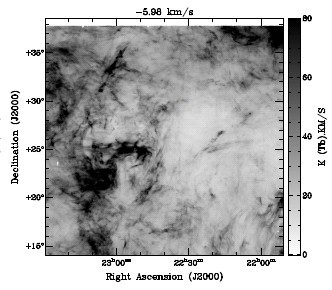}
\includegraphics[angle=0,width=0.33\textwidth,bb=0 0 240 207]{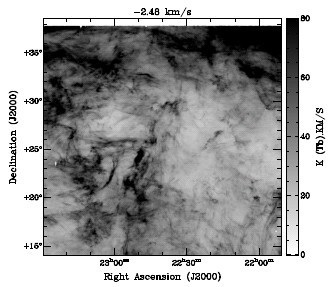}
\caption{HI channel maps of GSH 90-28-17 for the velocity range from $-26.4$~km~s$^{-1}$
    to $-2.5$~km~s$^{-1}$, averaged over 19 consecutive channels, yielding a velocity
    resolution of $\sim 3.4$~km~s$^{-1}$. The central LSR velocities are indicated on top of
    each panel. }
\label{chan}
\end{figure*}

In Fig.~\ref{chan} we show the HI-channel maps of GSH 90-28-17 in the
velocity range between $v=-27$ and $-6$~\kms. Each panel represents the 
integrated image over 19 consecutive channels, yielding a velocity resolution of
$\sim 3.4$~\kms. The central velocity of each panel is indicated at the top.
The filamentary shell structure is clearly visible with a large void in the
center of the images. The center of the shell is at
$l=89\degr, b=-31\degr, v=-17$~\kms, which is at a location slightly different
from that given in~\citet{h84}. The western shell mainly lies at $v\approx-21$~\kms,
while the eastern shell lies at $v\approx-12$~\kms. This may indicate that the
northwestern shell is on the far side, or expands backwards, while the southeastern
shell is on the near side, or approaching us.

\begin{figure}
\centering
\includegraphics[angle=0,width=0.42\textwidth,bb=0 0 328 232]{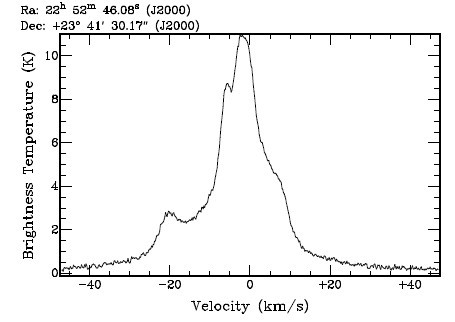}
\caption{HI velocity profile through the center of GSH 90-28-17, showing the dip
    of the shell at $v=-17$~km~s$^{-1}$ and the peaks at its front and back caps.
    The front cap at $v=-13$~km~s$^{-1}$ shows a bit of confusion with the local background
    emission.}
\label{line}
\end{figure}

Fig.~\ref{line} shows an HI velocity profile through the shell center.
Towards the high latitude direction the local background for the HI intensity is not strong.
The shell can be clearly identified in the profile as a dip at $v=-17$~\kms.
The back cap of the shell is apparent as a bump in the velocity profile
at $v\approx-21$~\kms on the left side of the void, while the front cap at
$v\approx-12$~\kms is a bit confused with the background emission.

\begin{table}
\caption{Basic parameters for GSH 90-28-17}
\label{para}
\begin{center}
\begin{tabular}{l l}
\hline\hline
 Parameter             & Value   \\\hline
 Center $l$            & 89\degr  \\
 Center $b$            & $-$31\degr \\
 Central $v_{\rm lsr}$ & $-$17~km~s$^{-1}$ \\
 Distance $d$          & 400~pc  \\
 Radius $r$            & 66.0$\pm$3.5~pc  \\
 Expansion velocity $v_{\rm exp}$     & 4.5~km~s$^{-1}$ \\
 Column density $N$    & (5.3$\pm0.2)\times 10^{20}$~cm$^{-2}$    \\
 HI gas Mass           & (3.1$\pm0.1)\times 10^{4}~M_{\odot}$  \\
 Kinetic energy $E_{kin}$   & (8.2$\pm0.3)\times 10^{48}$~ergs  \\

\hline
\end{tabular}
\end{center}
\end{table}

\begin{figure}
\centering
\includegraphics[angle=0,width=0.48\textwidth,bb=0 0 260 227]{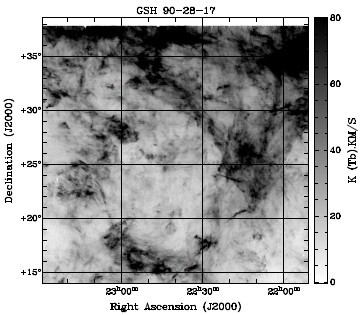}
\caption{HI intensity map of GSH 90-28-17, integrating from velocity $v=-28$ to
    $-15$~\kms. The shell is typically spherical with a diameter of about 20\degr.}
\label{gsh}
\end{figure}

In Fig.~\ref{gsh} we present the integrated HI intensity map of GSH 90-28-17, 
integrating over the velocity range $-28$ to $-15$~km~s$^{-1}$. 
The HI intensity get enhanced towards the north, which is the direction of the 
Galactic plane. The three-dimensional morphology of the shell is typically spherical, 
and extends over approximately 20\degr. Assuming that the HI emission is
optically thin, the HI column density is given by:
$N_{\rm HI}=1.8\times10^{18}\int T(v)dv$~cm$^{-2}$, where $T(v)$ is the brightness
temperature at LSR velocity with the background subtracted. We obtain the average
column density over the shell region as (5.3$\pm0.2)\times 10^{20}$~cm$^{-2}$.

\begin{figure}
\centering
\includegraphics[angle=0,width=0.42\textwidth,bb=0 0 318 236]{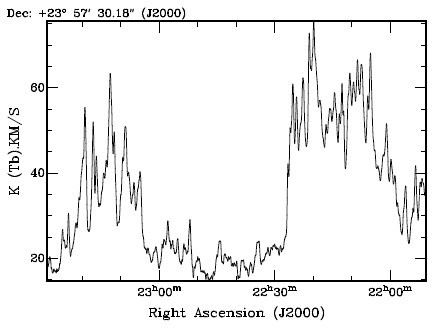}
\caption{Ra profile of the slice across GSH 90-28-17 at Dec$=+23\degr57\arcmin30.18\arcsec$ 
    of the integrated HI map, showing two sharp walls and a void in the shell center.
    The contrast factor in the brightness temperature between the walls and the 
    shell interior is 2$-$3, indicating the shell has a compression origin.}
\label{wall}
\end{figure}

With the sensitivity of GALFA, some patchy HI structures are clearly seen
inside the shell. The prominent one is the cloud in the interior of the shell
at (Ra, Dec, $v$)=(22$^{h}$45$^{m}$, +26$\degr$, $-$13~km~s$^{-1}$), at an average
level of $T_{b}\sim 15$~K.  In a largely evacuated area, the existence of 
such a gas clump is unusual. It is probably a pre-existing molecular cloud that
swept by the shocks. Figure~\ref{wall} shows a profile that is a slice across the shell
in the Ra direction. The slice is taken at Dec $+23\degr57\arcmin30.18\arcsec$
of the integrated HI map. The mean brightness temperature in the interior of
the shell is about $T_{b}=20$~K, shown as the low-level base structure. 
The walls show a brightness temperature contrast factor of $2-3$ with respect to the shell
interior over about 10 resolution grids. The sharpness of the wall indicates 
the compression is strong, probably associated with a shock.

\subsection{The distance}
Since many derived properties depend strongly on the adopted distance, an
accurate estimate of the distance is essential.

The techniques for distance estimates based on rotation velocity are not 
suited for shells at high latitude~\citep{h81}. We firstly follow the scale
height method in~\citet{h81} to re-estimate the distance of GSH 90-28-17 using
new parameters. The method sets the measured column density of the HI shell, 
$N_{HI}$, to be equal to the value one would expect to sweep up with a shell 
of the measured extent, $\Delta D$, at a distance of $d$,
as $N_{HI}=n(z)(d\Delta D)$. And it assumes the density distribution of material
follows $n(z)=n_{0}~exp(-|\frac{z}{h}|^{2})$, where $z=d~sinb$ is the height
from the Galactic plane. Taking a typical density $n_{0}=1$~cm$^{3}$, and 
scale height $h=$130~pc~\citep{s78}, the distance $d$ could be solved iteratively 
in combination with these two equations.
 
We solve the distance of GSH 90-28-17 with the parameters listed in Table~\ref{para}. 
However, the procedure does not produce a consistent solution. 
Such a situation also happened for many shells in~\citet{h81}, when the column density attributed to the shell is too large. The physical explanation is that the 
shells originate in regions of high density, and collect more material than 
that can be sweep up according to the assumption.

High latitude supershells are usually local gas structures. The
absorption associated with the shell of diffuse X-ray emission towards
GSH 90-28-17 at 0.25~keV also confirms it as a local shell (in
Sect.~3.7).~\citet{h81} gave a lower limit for the distance of 300~pc for the
partial shell GS~91$-$33 of GSH 90-28-17 based on extinction versus distance studies
along the line of sight. We note that the famous Lac OB1 association
(with center around RA=22h35m and Dec=+43.3\degr) lies in the upper left of
GSH 90-28-17 ~\citep{cl08}. It has an average distance of
$\sim$370~pc, which was derived from the $Hipparcos$ data~\citep{zhb99}. Lac OB1
is part of the local cloud related to the Gould belt, which constitutes
the expanding Lindblad ring in whose periphery lies the local
stellar associations~\citep{cl08}. The enhanced HI column density of
GSH 90-28-17 indicates that it probably also lies in the Gould belt, at
a distance similar to Lac OB1. The $Hipparcos$ catalog gave an average
radial velocity of Lac OB1 of $-13.3$~\kms~\citep{zhb99}, which is
quite close to that of GSH 90-28-17. Thus we suggest that GSH
90-28-17 is related to the Lac OB1 association. As the distance of nearby
association has a noticeable range of a few hundred parsecs, 
we adopt a distance of 400~pc for GSH 90-28-17 in the latter part of the paper.

\subsection{The Physical Properties}
The physical properties of GSH 90-28-17, such as expansion velocity
and kinetic energy, are re-estimated based on the new distance. 

Expansion velocities for shells are usually estimated as half of the total measured
velocity width, $\Delta v$, of the shell. The full velocity width of GSH 90-28-17,
through the center of the shell, is approximately $\Delta v=9$~km~s$^{-1}$. Thus
the expanding velocity is $v_{\rm exp}\approx4.5$~\kms.   

Based on the distance of 400~pc, the linear radius of GSH 90-28-17 is about
66.0$\pm$3.5~pc. The linear shell thickness is about 14~pc from the
apparent thickness of 2\degr. The HI mass ($m_{\rm H}N_{\rm HI}\times$ surface area) 
in the mapped region is (3.1$\pm0.1)\times 10^{4}~M_{\odot}$ (d$\sim$400~pc).  
The total mass of GSH 90-28-17 is $M_{tot}=1.3M_{HI}$, taking into account 
primordial helium. The kinetic energy associated with the expansion of the 
shell is estimated following the formula $E_{kin}=\frac{1}{2}M_{tot}^{ }v_{exp}^{2}$, 
which is (8.2$\pm0.3)\times10^{48}$~ergs.
The total energy of the shell which accounts for the observed radius and expansion
is a combination of the kinetic energy and thermal energy within the shell.
All these physical quantities are summarized in Table.~1.

We estimate the ambient density in the environment. If this mass of neutral gas
had originally been uniformly distributed within a sphere of radius 10\degr, 
the pre-explosion ambient density would have been 0.9~cm$^{3}$. 
The value is consistent with that derived from the averaged background column
density in the vicinity of GSH 90-28-17.

The derived ages of HI shells are usually associated with large
uncertainties. Accurate estimation requires a powering source
associated with the shell, whose age should be independently measured. 
However, we can estimate the dynamic age based
on models of a shell expanding from the pressure of a hot interior.
For GSH 90-28-17, the shock radius follows an expansion law $R\propto
t^{0.3}$~\citep{cmb88}, assuming an evolution path similar to that of supernova
remnants in the late radioactive phase.  Hence the dynamic age $t_{6}$ in
units of Myr is $t_{6}=0.29R/v_{exp}$, where the radius $R$ is in pc
and $v_{exp}$ is in ~\kms.  The resulting dynamic age of GSH 90-28-17
is $\sim 4.5$~Myr.  It is relatively old compared with other known
Galactic shells~\citep{mdg02}.

In Sect.~3.2, we suggest the shell is associated with the Lac OB1 association.
Lac OB1 has an expansion time scale of tens of Myr, on the basis of stellar proper
motions and radial velocities~\citep{cl08}. It is in its final stage of star formation,
and the last star formation episode took place no more than a few Myr ago.
The main sequence lifetime of the only O star in Lac OB1 is 3.6~Myr~\citep{sd97}.
Based on the age estimation, the star that is the energy source of GSH~90-28-17 is likely
to be formed earlier than the last star formation in the association, {\it i.e. }
has an age of more than a few Myr.

\begin{figure*}
\centering
\includegraphics[angle=0,width=0.48\textwidth,bb=0 0 267 209]{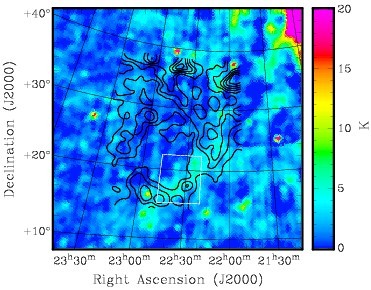}
\includegraphics[angle=0,width=0.48\textwidth,bb=0 0 246 193]{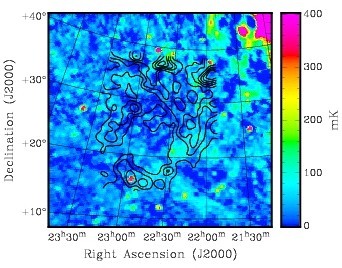}
\caption{The filtered 408~MHz and 1420~MHz continuum image obtained by applying the
    ``background filtering'' (BGF) method. The contours show the integrated HI 
    map smoothed to 1\degr, which start at 22.5~K~\kms and increase in steps of
    15~K~\kms. The box marks the TT-plot region.}
\label{bgf}
\end{figure*}

\subsection{Radio continuum}
We extracted the area of GSH 90-28-17 from the 408~MHz all-sky Survey and Bonn 1420~MHz
survey data~\citep{hss82,r82}. The diffuse Galactic background continuum emission
decreases with distance from the Galactic plane. A region with weak continuum emission
is associated with GSH 90-28-17, with a continuum ridge stretching out over the Galactic
plane overlapping with the western shell.

In order to reveal radio emission overlapped with the HI shell, we
subtracted the smooth Galactic background emission following the same
method as~\citet{sn83}.  We applied the ``background filtering'' (BGF)
procedure described by~\citet{sr79} to the original radio data. The
filtering beam used is about three times the original resolution,
2.5$\degr$ with cut-values of 50~K~T$_{\rm B}$ at 408~MHz and 2$\degr$
with 5000~mK~T$_{\rm B}$ at 1420~MHz, respectively.  Structures with
scale smaller than the filter beam will stand out in the filtered
image.  Different filtering beam widths (2-3\degr) essentially show
the same features, except that a smaller beam width will slightly
reduce the intensity.  The filtered maps obtained by applying the BGF
method are present in Fig.~\ref{bgf}. A prominent radio loop-like
ridge with a length of approximately 10\degr~is clearly revealed to be
associated with the western HI shell at both frequencies. A radio
point-like source with peak temperatures of 680~mK at 1420~MHz seems
coincident with an HI enhancement.  The excess brightness of the
loop-like ridge over the background is 3$-$6~K at 408~MHz, and
60$-$100~mK at 1420~MHz. The rms fluctuations $\Delta T_{b}$ in the
residual images of 408~MHz and 1420~MHz in Fig.~\ref{bgf} are 0.4~K
and 10.0~mK, respectively.

\begin{figure}
\centering
\includegraphics[angle=0,width=0.38\textwidth,bb=0 0 274 176]{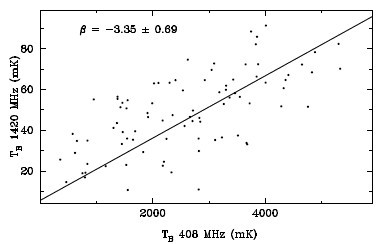}
\caption{The TT-plot of the radio loop-like emission associated with the western shell
     of GSH~90-28-17. Both residual maps after BGF filtering at 408 MHz and 1420 MHz
     are smoothed to a common resolution of 1\degr, and the grids are re-sampled to 
     one point within one beam. The fitted spectral index is $\alpha$=-1.35$\pm$0.69.}
\label{tt}
\end{figure}

Although the BGF technique is effective for demonstrating faint features,
the brightness temperatures on the resulting map usually give
underestimated values.  In order to investigate the spectral index
properties of the radio loop-like emission, we applied the TT-plot, which
is unaffected by the uncertainty of the baselevels, at these two
frequencies. Both residual maps after BGF filtering at 408~MHz and
1420~MHz are smoothed to a common resolution of 1\degr before calculating the
spectral index. And the grids are re-sampled to one point within one beam
to avoid the oversampling.
The TT-plot region is marked in Fig.~\ref{bgf}.
The resultant TT-plot for the western radio loop-like
emission is shown in Fig.~\ref{tt}. It has a steep spectral index of
$\alpha$=-1.35$\pm$0.69 ($S_{\nu}\propto\nu^{\alpha}$, $\alpha=\beta+2$). 
The error is obtained by fitting the data twice, alternatively taking the 
data of one of the two wavelengths as the independent variable.
The large error of the TT-plot is dominated from
the uncertainties of background emission level. However, 
the steep spectral index indicates a possible nonthermal origin of the radio emission, 
and it is consistent with the expectation of synchrotron emission from old SNRs.
The spectral index in an old SNR probably becomes steeper due to
synchrotron losses of electrons. One example is SNR S147, for which a
spectral break was found in its spectrum and the diffuse emission
component has a steep index of $\alpha\sim-1.35$~\citep{xfr08}.

Unlike the radio emission from normal SNRs which is usually distributed
within the HI shell, the radio loop-like emission is somewhat overlapped with 
GSH 90-28-17. Detailed confirmation of the nonthermal property of
the radio loop-like emission and the association with GSH 90-28-17
requires new high-resolution radio observations towards this direction. 
Assuming there is a physical relation between them, we could
estimate the radio surface brightness for GSH 90-28-17.
We integrated over the radio loop associated with the western HI shell, and got
flux densities of 322.0~Jy and 68.3~Jy at 408~MHz and 1420~MHz, respectively.
The errors are not estimated for the residual maps after filtering.
If we take into account the upper limit of the brightness temperature $T_{b}$ at
other regions of the shell, we can obtain a total flux density of $\sim$514.0~Jy for 
GSH 90-28-17 at 408~MHz. Extrapolating it to 1~GHz using the spectral index of $-$1.2,
we estimate the surface brightness of
$\Sigma_{1GHz}$ as $ < 5.1\times10^{-23}$~Wm$^{-2}$Hz$^{-1}$sr$^{-1}$. 
This value is comparable to that of the weakest SNR in our Galaxy of
SNR G156.2+5.7~\citep{xhs07} and G65.2+5.7～\citep{xrf09}.

\begin{figure}
\centering
\includegraphics[angle=0,width=0.48\textwidth,bb=0 0 240 209]{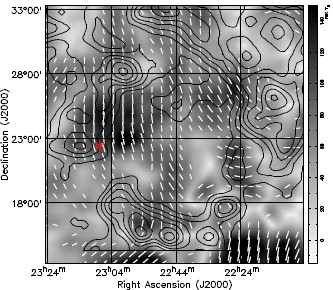}
\caption{DRAO 21~cm polarized intensity (PI) map of GSH 90-28-17 at an angular
    resolution of 36\arcmin. The bars indicate the direction of the magnetic field, with
    lengths proportional to PI. Contours show the integrated HI map, starting at
    22.5 K~\kms and increasing in steps of 7.5 K~\kms. The position of the pulsar J2307+2225 is
    indicated as an asterisk.}
 \label{pi}
\end{figure}

\subsection{Polarization}
We extracted the 21~cm polarized intensity map of the GSH 90-28-17 region
(Fig.~\ref{pi}) from the polarization survey made with the DRAO 26~m telescope
~\citep{wlr06}. This map includes the large-scale polarized emission component
at an absolute zero level with an angular resolution of 36\arcmin. 
We found no prominent polarization emission associated with the radio loop-like ridge,
except for some polarization patches in the shell region with randomly distributed
polarization angles. This evidence does not seem to support the SNR origin of the shell. 
However, the magnetic field in the radio ridge might be weakened as the SNR evolves into
the late stage. And it is also possible that the radio ridge depolarized the emission
from larger distances.

\subsection{Pulsars}
We searched through the ATNF pulsar catalog to find possible associated pulsars.
Three pulsars (J2234+2114, J2229+2643 and J2307+2225) are located within the HI
shell. J2229+2643 has a short period of $P \sim 0.003$~s, and seems to be a
millisecond pulsar which was formed by accretion. It could be excluded from the candidates
for possible association with the HI shell. The periods of J2234+2114 and J2307+2225 are
1.4 and 0.5~s, with distances of 3.38 and 0.38~kpc, respectively,
from the Dispersion Measure based on the Galactic electron density model~\citep{tc93}.
Although the characteristic age of the pulsar J2307+2225 is about $9.8\times10^{8}$~yr
~\citep{cn95}, derived from the ratio $p/2\dot{p}$ (p is the period of the pulsar),
the true age of the visible pulsar should be younger than that from this linear derivation, 
and rarely exceeds some $10^{6}$~yr~\citep{s77}. It appears that J2307+2225 is the most probable
candidate pulsar, and its position is marked in Fig.~\ref{pi}.
If J2307+2225 is associated with GSH90-28-17, we can estimate the kick velocity
of the pulsar from its current location relative to the center of the shell,
which is about 14~\kms.

\subsection{Comparison with Other Wavelengths}
\begin{figure*}
\centering
\includegraphics[angle=0,width=0.49\textwidth,bb=0 0 238 202]{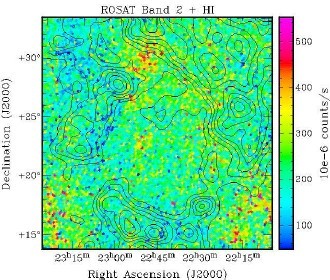}
\includegraphics[angle=0,width=0.49\textwidth,bb=0 0 235 201]{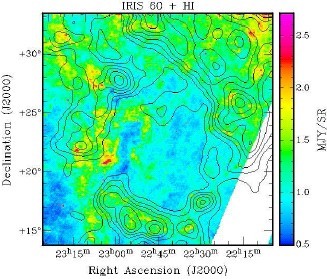}
\caption{The X-ray Band 2 image and $IRIS$ 60~$\mu$m dust emission of GSH 90-28-17
     overlaid with contours of the integrated HI map smoothed to 1\degr. 
     The contours start from 22.5 K~\kms and increase in steps of 7.5 K~\kms.}
\label{multi}
\end{figure*}

We check the diffuse X-ray emission towards GSH90-28-17, using the publicly
available data from the $ROSAT$ all-sky background survey~\citep{sef97}.
The 0.25, 0.75 and 1.5~keV emission is compared with the HI distribution. 
There is clear evidence of 0.25~keV absorption associated with the shell region. 
The RASS Band 2 (0.14$-$0.284~keV) image of GSH90-28-17 overlaid with contours
of HI integrated intensity is shown in Fig.~\ref{multi}.
At 0.25~keV, one optical depth corresponds to HI column densities of about
$1\times10^{20}$~cm$^{-2}$~\citep{sef97}, which is lower than the averaged
column density in the HI shell region. Considering an averaged space density
of 0.1~cm$^{-3}$ (averaged over the local bubble) along the line of sight, 
the mean free path for 0.25~keV X-rays is about $\sim324$~pc. 
Thus most of the R2 band emission must have originated locally. 
This further excludes the possibility of a farther distance for GSH90-28-17.
The absorption become less prominent at 0.75~keV, and is not seen 
at 1.5~keV. The mean free paths for X-rays in the higher energy bands are much longer.
At the 1.5~keV band, one optical depth is $\sim5\times10^{21}$~H~cm$^{-2}$~\citep{sef97}.

However, we find no obvious X-ray excess that traces the interior hot gas in
all these three bands. It is not unexpected for the lack of hard X-ray features, 
because 0.75~keV and 1.5~keV emissions require very hot gas with temperatures 
on the order of $10^{7}$~K. Old large shells, such as GSH 90-28-17, are not 
expected to have gas temperature much higher than about 
$3.5\times10^{6}$~K~\citep{mm88}. At 0.25~keV, most of the emission originates
from thermal radiation from a 10$^{6}$~K component of the interstellar gas within
a few hundred parsecs of the Sun~\citep{ms90}, and the excess from GSH 90-28-17 
is not significantly larger.

We have also compared the HI distribution of GSH 90-28-17 with the H$\alpha$
emission from the H$\alpha$ composite survey~\citep{f03}. 
H$\alpha$ in the inner wall of the shell is weak, indicating
that the temperature and/or electron density there is low.
The $IRIS$ 60~$\mu$m dust emission with HI intensity contours superposed is
shown in Fig.~\ref{multi}. There is dust emission associated with the shell region,
especially in the eastern shell.

\section{Discussion}

Both the energy released in a stellar wind and a supernova explosion could create
an HI shell structure with size  similar to that of GSH 90-28-17. Although the
pulsar J2307+2225 detected within the HI shell and the association with non-thermal radio
continuum emission at 408 and 1420~MHz seem to favor a supernova event.
The expansion kinetic energy of GSH 90-28-17 is comparable with another HI shell
G132.6-0.7-25.3~\citep{ntd00}. The size of GSH 90-28-17 is about twice that of G132.6-0.7-25.3 (radius 33~pc), which is probably caused by the relatively low ambient density. ~\citet{ntd00} found that the B1 Ia star BD +60\degr447 could have
been the energy source of the shell by means of its stellar wind, but they also
could not exclude the possibility of an old supernova remnant.

In this section, we shall explore the possibility of stellar wind and supernova
formation scenarios of GSH 90-28-17 based on their theoretical models, individually. 
If the observed HI shell is the result of only one stellar wind, what would
be the source star of GSH 90-28-17? As a less massive progenitor star would be required
in the supernova case, then which type of supernova it would be?

\subsection{Stellar wind scenario}

The stellar wind of a single O-type or an early B-type star would be strong enough
to blow a bubble while on the main sequence, and then no longer be capable of
maintaining the ionization of the shell. The energy injected by the stellar winds
$E_{w}=\dot{M}v_{w}^{2}t/2$ is related to the luminosity of the star. 
Thus we can estimate the luminosity of the star required to blow the
kinetic energy of GSH 90-28-17.

The relation between mass-loss rate $\dot{M}$ and the stellar luminosity
is obtained from Table.~1 of~\citet{lhk99} for a sample of O and B stars.
The wind velocity $v_{w}$ is also related to the stellar luminosity at an average
level of about 2000~\kms. The main-sequence lifetime of stars with different
luminosity could be estimated through the stellar evolution models for a solar
metallicity of~\citet{bfb93}.

When the interstellar bubble expands, only a fraction of the stellar wind 
energy $\epsilon E_{w}$ is transferred to the kinetic energy of the HI gas. 
Some fraction is converted into thermal energy when the stellar wind shock 
sweeps up the interstellar medium. The expected energy conversion efficiency 
$\epsilon$ is on the order of 0.2 or less~\citep{m83}, and sometimes can be 
as low as 0.02~\citep{cac03} when severe energetic losses occur.

If we adopt a typical conversion efficiency of 0.2, the expansion energy of
GSH 90-28-17 requires a star with luminosity log($L/L_{\odot}$)=5.3 at solar
metallicity. This luminosity implies there is a main-sequence star with mass
of 30~M$_{\odot}$, lifetime of 6.2~Myr~\citep{bfb93}, and spectral type of 
approximately O7V. The age of GSH 90-28-17 estimated from the expansion law 
is $\sim$4.5~Myr, similar to the main-sequence lifetime of a 30~M$_{\odot}$ star.

The distance of GSH 90-28-17 is local. If the star still exists, it should be 
detectable. Based on the star evolutionary tracks~\citep{bfb93},
and using the aforementioned luminosity and effective temperature,
the star should now appear as a B1 star. However, there are very few massive stars
in the center of the shell. As we searched the SIMBAD database, there are only 12 B
stars, of which the earliest is a B8-type star, which is located near/in the shell walls. 
The central star is a B8 star, with a distance of $\sim$870~pc,
which is too far to be associated with the HI shell. Thus the stellar wind 
(blown-bubble) scenario seems to be difficult for explaining the origin of the shell.

\subsection{A supernova explosion}
The association between non-thermal radio continuum emission at 408 and 1420~MHz with
GSH 90-28-17, and the pulsar J2307+2225 detected within the HI shell indicates
that G90-28-17 is probably an old supernova remnant.

After the cooling-dominated radiactive expansion phase, old SNRs will evolve into a
late stage, and eventually merge with the interstellar medium and become indistinguishable. 
Adopting a distance of 400~pc, we extrapolate the initial
explosion energy of GSH 90-28-17 from the maximum observable radius and the ambient
environment following the evolution model for a late stage SNR~\citep{cmb88,tgj98}, 
\begin{equation}\label{energy}
R_{\rm merge}=51.3E_{0}^{31/98}n_{0}^{-18/49}Z^{-5/98}
\end{equation}
where $R_{\rm merge}$ is the maximum observable radius of the shell in units of pc,
$E_{0}$ is the explosion energy in units of 10$^{51}$~ergs, $n_{0}$ is the
mean ambient particle density of 0.9~cm$^{-3}$, and $Z$ is the metallicity
normalized to the solar value. Adopting a metallicity $Z=1$, the energy required
to generate a shell by a single SN explosion is estimated to be about 2.3$\times10^{51}$~ergs,
which is the typical explosion energy output of a type II supernova.

The size and age of GSH 90-28-17 is comparable to supernova remnants that are
considered to be very old in the Galaxy. Good examples of such type are the
Monoceros SNR (G205.5+0.5) and radio Loop I (North Polar Spur). 
The age of the Monoceros SNR is 1.5$\times10^{6}$~yr,
with a radius of 53~pc and expansion velocity of about 8~\kms~\citep{xz12}.
The radio Loop I is considered to be two evolved supernova remnants
(S1 and S2) by~\citet{w07} based on the polarization observations at 21~cm. 
The modeled radius of these two supernova remnants are about 75~pc (S2) and 
81~pc (S1), with age $1-2\times10^{6}$ and $6\times10^{6}$~yr, respectively.
The expansion velocity of the associated HI shells has two values: 2~\kms
from the northern hemisphere (GS 331+14-15) measured by~\citet{w79} and 19$-$25~\kms from
the southern hemisphere measured by~\citep{h84}, corresponding to these two
supernova remnants that are expanding with different velocities. 
The properties of the S1 supernova remnant are very
similar to those of GSH 90-28-17. This shell is almost dissolved with no observational
tracers (except for the polarization emission loop), and could not be found in 
total intensity surveys.

The SNR interpretation of GSH 90-28-17 makes it among the oldest SNRs known.
Our case study of this source also demonstrates the capability of high-resolution HI surveys in finding the oldest supernova remnants. Though the radio and X-ray
emission in the center of the remnant is rather weak, we are still able to see 
the well-defined HI shell of GSH 90-28-17. It is generally assumed that the 
accumulated expanding gas shell will survive 10 times longer until the expansion 
velocity decreases down to the random velocity of the interstellar medium~\citep{mo77}, 
and then dissolves into the ambient medium.
As the shell remains as a high-density enhancement in a low-density region 
for a long time, it provides a good environment to allow star formation to progress after
a single trigger event. ~\citet{si01} have interpreted another HI supershell GSH 138-01-94 as
an old SNR with a similar age of 4.3~Myr in the far outer Galaxy.
It was later suggested by~\citet{kyt08} to have triggered the star formation 
in Digel's cloud 2. As a high latitude supershell far from the Galactic plane,
GSH 90-28-17 greatly avoids the absorption effect caused by intervening material, 
and provides a good example to study the last phase of evolution for old SNRs in 
a relatively peaceful environment of the Lac OB1 association.

\section{Conclusions}

The properties of the HI supershell GSH 90-28-17 are re-examined using the new Arecibo
GALFA-HI survey data with greatly improved resolution and sensitivity. 
We suggest that GSH 90-28-17 is related to the Lac OB1 association, with
a distance of 400~pc. The radius of GSH 90-28-17 is 66.0$\pm$3.5~pc. The HI mass
of the shell is about (3.1$\pm0.1)\times10^{4}$~M$_{\odot}$.
It has an age of $\sim4.5$~Myr and a total kinetic energy of about 
(8.2$\pm0.3)\times10^{48}$~ergs.

We used the filtering algorithm (BGF) following~\citet{sn83} to subtract the 
large-scale diffuse Galactic background emission from 408 and 1420~MHz survey maps, 
and found a radio loop-like ridge associated with the western shell of GSH 90-28-17
at both frequencies. The excess brightness of the radio ridge over the background 
is $3-6$~K at 408~MHz and $60-100$~mK at 1420~MHz. The TT-plot index of the radio 
loop-like ridge is $\alpha$=-1.35$\pm$0.69, shows a nonthermal origin. In addition, 
the pulsar J2307+2225 is found inside the shell region at a similar distance.

We discussed the possibility of stellar wind and SNR scenarios for the GSH 90-28-17.
If it has an origin from stellar wind, the energy source would be a 30~M$_{\odot}$, 
O7 star, and should now be observable as a B1 star. However no star earlier 
than B8 is found within the shell. We conclude that GSH 90-28-17 is probably 
an old, type II SNR in the Galaxy around the solar neighborhood.

\section*{Acknowledgments}

The GALFA-HI survey was obtained with a 7-beam feed array at the
Arecibo Observatory. We are grateful to the staff at the Arecibo
Observatory, and the GALFA-HI survey team for conducting the GALFA-HI
observations. This work is partly supported by the China Ministry of
Science and Technology under State Key Development Program for Basic
Research (2012CB821800) and partly supported by the General Program of
National Natural Science Foundation of China (11073028, Y311431001,
Y123031001). We also thank the support from the Hundred
Talents Program of the Chinese Academy of Sciences, and thank 
Dr. James Wicker for proofreading the manuscript.

\end{document}